\documentclass[fleqn,twoside]{article}
\usepackage{espcrc2}

\usepackage{graphicx}
\usepackage[figuresright]{rotating}

\newcommand{\AmS}{{\protect\the\textfont2 A\kern-.1667em\lower.5ex\hbox{M}\kern-.125emS}}
\newcommand{\beqa}{\begin{eqnarray}}
\newcommand{\eeqa}{\end{eqnarray}}
\newcommand{\beq}{\begin{equation}}
\newcommand{\eeq}{\end{equation}}

\title{
{\noindent\small 
UNITU-THEP-7/04 \hspace*{1cm} IPPP/04/34  \hspace*{1cm} DCPT/04/68 \hfill hep-ph/0407294 }\\
Dynamical Mass Generation in Landau gauge QCD 
\thanks{Summary of a talk given at the international conference QCD DOWN UNDER, March 10 - 19, Adelaide, Australia}
	}

\author{C.~S.~Fischer\address{IPPP, University of Durham, Durham DH1 3LE, U.K.},{}
        F.~Llanes-Estrada\address{Fisica Teorica I, Univ. Complutense, Madrid 28040, Spain},{}
        R.~Alkofer\address{Institute for Theoretical Physics, University of
       T\"ubingen, D-72076 T\"ubingen, Germany}}
       
\begin{document}

\begin{abstract}
We summarise results on the infrared behaviour of Landau gauge QCD from 
the Green's functions approach and lattice calculations. Approximate, 
nonperturbative solutions for the ghost, gluon and quark propagators as 
well as first results for the quark-gluon vertex from a coupled set of 
Dyson-Schwinger equations are compared to quenched and unquenched lattice 
results. Almost quantitative agreement is found for all three propagators.
Similar effects of unquenching are found in both approaches. The dynamically
generated quark masses are close to `phenomenological' values. First results 
for the quark-gluon vertex indicate 
a complex tensor structure of the non-perturbative quark-gluon interaction.
\end{abstract}

\maketitle

\section{Introduction}
The nonperturbative properties of the quark-gluon interaction are at the heart 
of the most interesting phenomena of QCD such as confinement and dynamical chiral 
symmetry breaking. On a fundamental level the details of the quark-gluon interaction
are encoded in the one-particle irreducible quark-gluon vertex. Together with
the quark and gluon propagators this vertex enters as a vital ingredient in model 
building and plays a key role in bridging the gap between the coloured fundamental 
degrees of freedom of the theory and the observed colour-neutral hadron states. 
On a phenomenological level the quark-gluon interaction manifests itself in the 
details of the quark-(anti-)quark potential. Relativistic effects play an important 
role in potential models and the proper choice of the Lorentz structure of the 
quark-antiquark interaction is vital to guarantee the agreement of the theoretical
predictions with the experimental data. Ultimately,
a derivation of these structures from the underlying quark-gluon interaction is
mandatory. To this end one has to know the properties of the most fundamental
Green's functions of QCD, the dressed propagator and vertex functions.

The infrared behaviour of the propagators of Landau gauge QCD has been investigated
extensively over the past years in both, lattice Monte Carlo simulations and the
continuum Green's functions approach. Lattice simulations are the only {\em ab initio} 
method known so far and are by now precise enough to pin down these propagators 
accurately in a large momentum range centered around 1 GeV. In the deep infrared,
however, lattice results are inevitably plagued by finite volume effects. In the 
continuum formulation of QCD the Dyson-Schwinger equations (DSEs) provide a tool 
complementary to lattice simulations. They can be solved analytically in the 
infrared. Furthermore numerical solutions over the whole momentum range are 
available by now. The truncation assumptions necessary to close the DSEs can be
checked in the momentum regions
where lattice results are available. In general
results from DSEs have the potential to provide a successful description of hadrons 
in terms of quarks and gluons, see \cite{Maris:2003vk,Alkofer:2000wg,Roberts:2000aa} 
and references therein.

\newpage

\section{Landau gauge QCD}

The ghost, gluon and quark propagators, $D_G(p)$, $D_{\mu \nu}(p)$ and $S(p)$, 
in Euclidean momentum space can be generically written as
\begin{eqnarray}
  D_G(p,\mu^2) &=& - \frac{G(p^2,\mu^2)}{p^2} \,,
  \label{ghost_prop}\\
  D_{\mu \nu}(p,\mu^2) &=& \left(\delta_{\mu \nu} - \frac{p_\mu
      p_\nu}{p^2} \right) \frac{Z(p^2,\mu^2)}{p^2} \, ,
  \label{gluon_prop} \\
  S(p,\mu^2) &=& \frac{1}{-i  p\!\!\!/\, A(p^2,\mu^2) + B(p^2,\mu^2)}\nonumber\\
  &=&  \frac{Z_Q(p^2,\mu^2)}{-ip\hspace{-.5em}/\hspace{.15em}+M(p^2)}
  \, .
  \label{quark_prop}
\end{eqnarray}
Here $\mu^2$ denotes the renormalisation scale and $G(p^2,\mu^2)$ and 
$Z(p^2,\mu^2)$ are the ghost and gluon dressing functions. The Dyson-Schwinger 
equations for these dressing functions have been solved in their continuum
formulation \cite{vonSmekal:1997is,Fischer:2002hn} as well as on a torus, i.e.
employing periodic boundary conditions \cite{Fischer:2002eq}. In the
continuum they can be solved analytically in the infrared and one finds 
simple power laws,
\begin{eqnarray}
  Z(p^2,\mu^2) &\sim& (p^2/\mu^2)^{2\kappa}, \nonumber\\
  G(p^2,\mu^2) &\sim& (p^2/\mu^2)^{-\kappa},
  \label{g-power}
\end{eqnarray}
for the gluon and ghost dressing function with exponents related to
each other. Hereby $\kappa$ is an irrational number,
$\kappa = (93 - \sqrt{2101})/98 \approx 0.595$ which depends slightly on the 
truncation scheme \cite{Lerche:2002ep,Zwanziger:2001kw}. Note that $\kappa>0$,
also shown in \cite{Watson:2001yv}, implies the dominance of the 'geometric' 
ghost degrees of freedom in the infrared in agreement with the Kugo-Ojima 
confinement criterion and Zwanziger's horizon condition 
\cite{Nakanishi:qm,Kugo:1979gm,Zwanziger:2003cf}. Recently these results from 
the Dyson-Schwinger approach have been confirmed independently in studies
of the exact renormalisation group equation \cite{Pawlowski:2003hq}.

A combination of the ghost and gluon dressing functions can be used to define 
the nonperturbative running coupling \cite{vonSmekal:1997is}
\beq
\alpha(p^2) = \alpha(\mu^2) \: G^2(p^2,\mu^2) \: Z(p^2,\mu^2).
\eeq
No vertex function appears in this definition; a fact that can be traced back 
to the ultraviolet finiteness of the ghost-gluon vertex in Landau gauge.
Note that the right hand side of this equation is a renormalisation group 
invariant, i.e. $\alpha(p^2)$ does not depend on the renormalisation point.
A further renormalisation  group invariant is the quark mass function, given by 
$M(p^2)=B(p^2,\mu^2)/A(p^2,\mu^2)$. From the power laws (\ref{g-power}) one finds
that the coupling has a fixed point in the infrared given by\\
$$
\alpha(0) = \frac{4 \pi}{6N_c}
\frac{\Gamma(3-2\kappa)\Gamma(3+\kappa)\Gamma(1+\kappa)}{\Gamma^2(2-\kappa)
\Gamma(2\kappa)} \approx 2.972 \\
$$
for the gauge group SU(3). The infrared dominance of the ghosts can be used to 
show that $\alpha(0)$ depends only weakly on the dressing of the ghost-gluon 
vertex and not at all on other vertex functions \cite{Lerche:2002ep}.

\begin{figure}[t]
\includegraphics[width=\columnwidth]{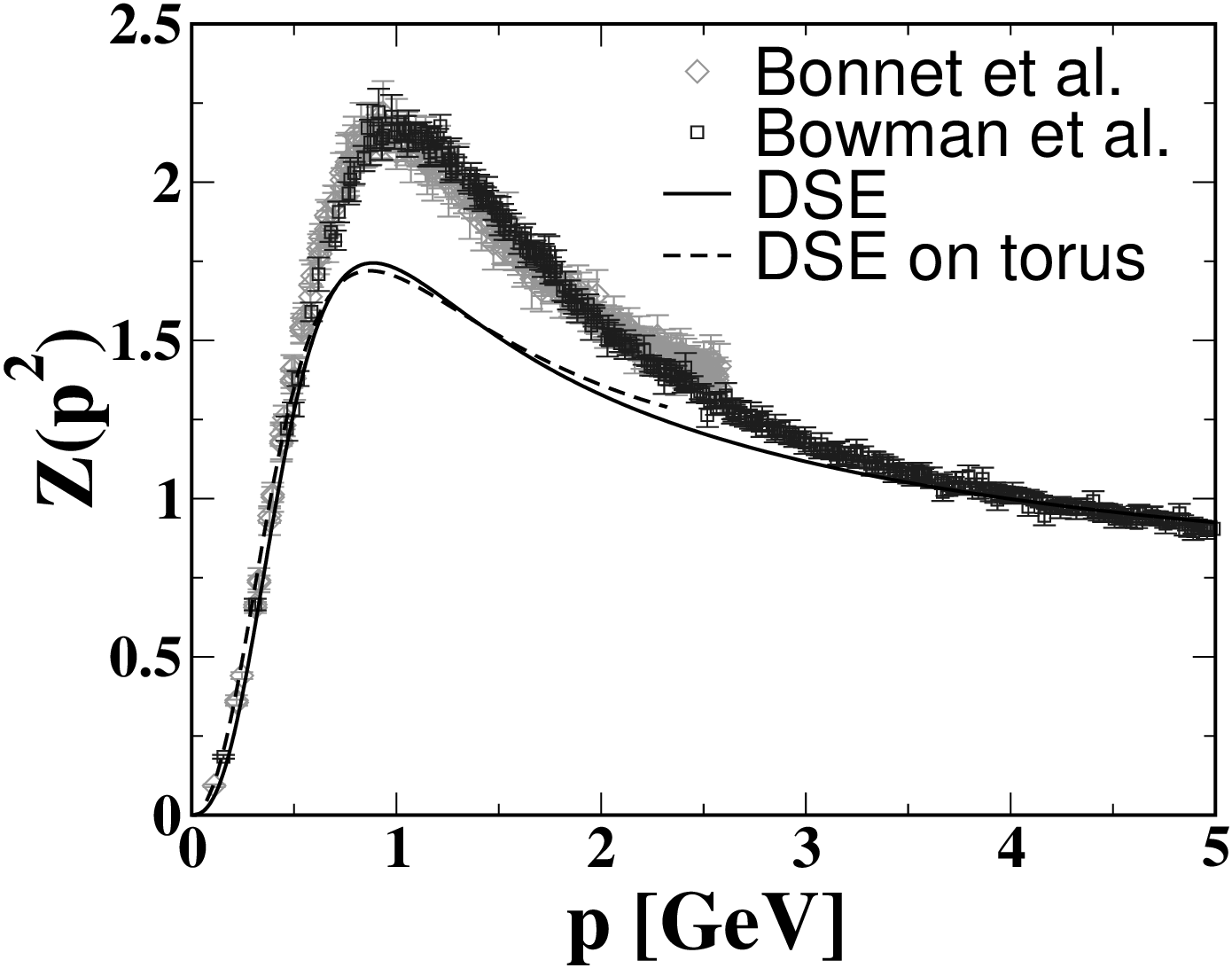}
\vspace{-1cm}
\caption{The gluon dressing function $Z(p^2)$ from DSEs 
\cite{Fischer:2002hn,Fischer:2002eq} and on the lattice 
\cite{Bonnet:2001uh,Bowman:2004jm}.} \label{fig1}
\end{figure}

The running coupling as it results from numerical solutions for the gluon,
ghost and quark propagators can be quite accurately fitted by 
the relatively simple function \cite{Fischer:2003rp}
\begin{eqnarray}
\alpha_{\rm fit}(p^2) &=& \frac{\alpha(0)}{1+p^2/\Lambda^2_{\tt QCD}}
+ \frac{4 \pi}{\beta_0} \frac{p^2/\Lambda^2_{\tt QCD}}{1+p^2/\Lambda^2_{\tt QCD}}
\times
\nonumber\\
&&\left(\frac{1}{\ln(p^2/\Lambda^2_{\tt QCD})}
- \frac{1}{p^2/\Lambda_{\tt QCD}^2 -1}\right) 
\label{fitB}
\end{eqnarray}
with $\beta_0=(11N_c-2N_f)/3$.
Note that, following ref.\ \cite{Shirkov:1997wi}, 
the Landau pole has been subtracted. 
The scale $\Lambda_{\tt QCD}$ is hereby determined
by fixing the running coupling at a certain scale, {\it e.g.\/}
$\alpha_S (M_Z^2) = 0.118$.

\begin{figure}[t]
\includegraphics[width=\columnwidth]{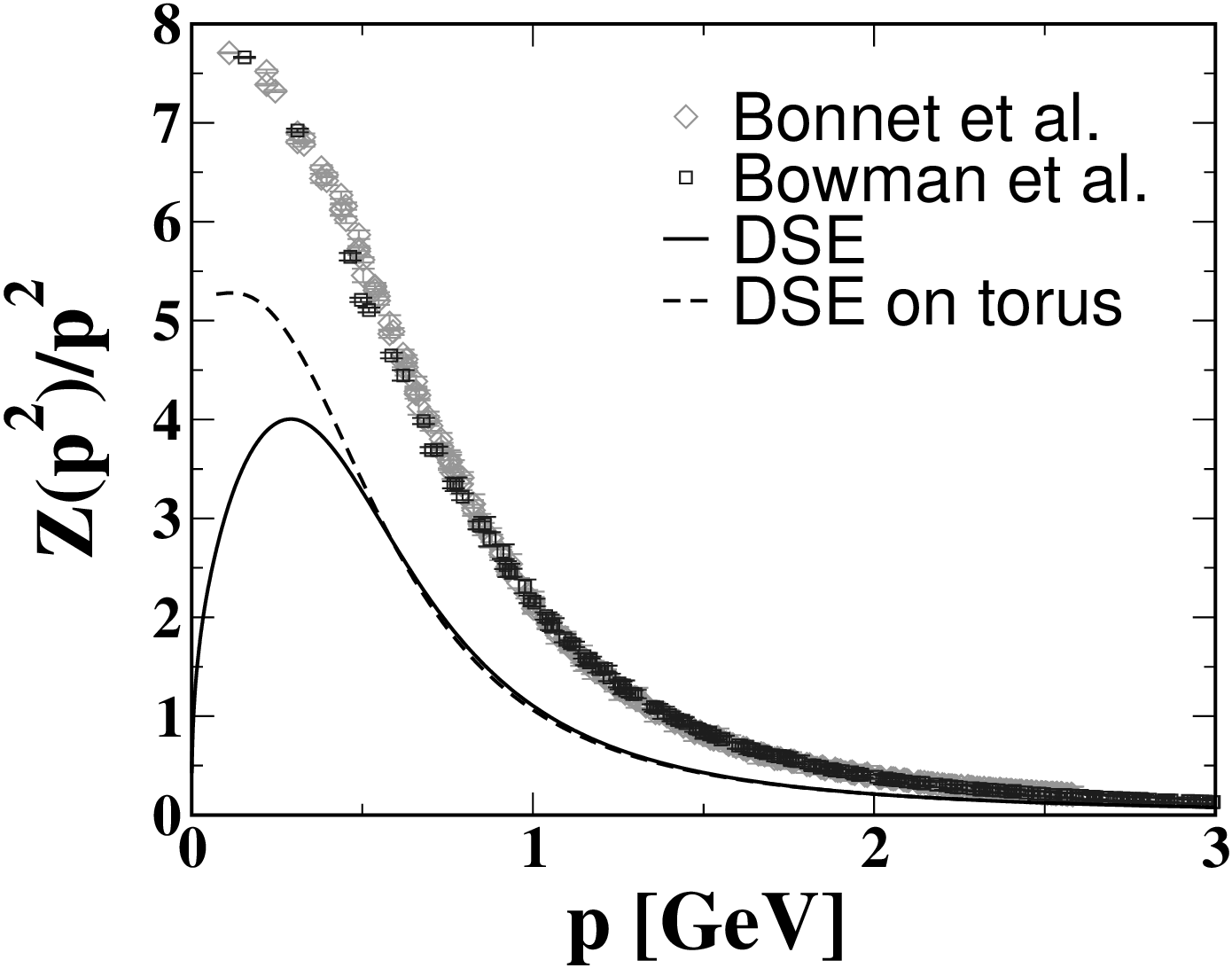}
\vspace{-1cm}
\caption{The gluon propagator $Z(p^2)/p^2$ 
from DSEs 
\cite{Fischer:2002hn,Fischer:2002eq} and on the lattice 
\cite{Bonnet:2001uh,Bowman:2004jm}.}\label{fig2}
\end{figure}

The numerical solutions for the gluon dressing function $Z(p^2)$ in the continuum 
and on the torus are compared to the results of recent lattice simulations 
\cite{Bonnet:2001uh,Bowman:2004jm} in fig. \ref{fig1}. The qualitative 
agreement with the lattice results is very good and the quantitative 
discrepancies in the momentum region around 1 GeV well understood: gluon
self interactions not contained in the DSE-truncation play an important 
role here. In the infrared both approaches agree nicely. An interesting
qualitative difference, however, can be seen in fig. \ref{fig2}, where we show
the propagator $Z(p^2)/p^2$ instead of the dressing function. Both, the
lattice results and the DSE-solution on the compact manifold tend towards a
constant in the infrared, whereas the propagator from the DSEs in continuum 
formulation vanishes. Both DSE-solutions employ the same truncation scheme,
thus the differences between them have to be attributed to the periodic 
boundary conditions and the finite volume on the torus.
These properties of the compact space might also account for the
constant gluon propagator found on the lattice.

\section{Dynamical mass generation}

\begin{figure}[t!]
\includegraphics[width=\columnwidth]{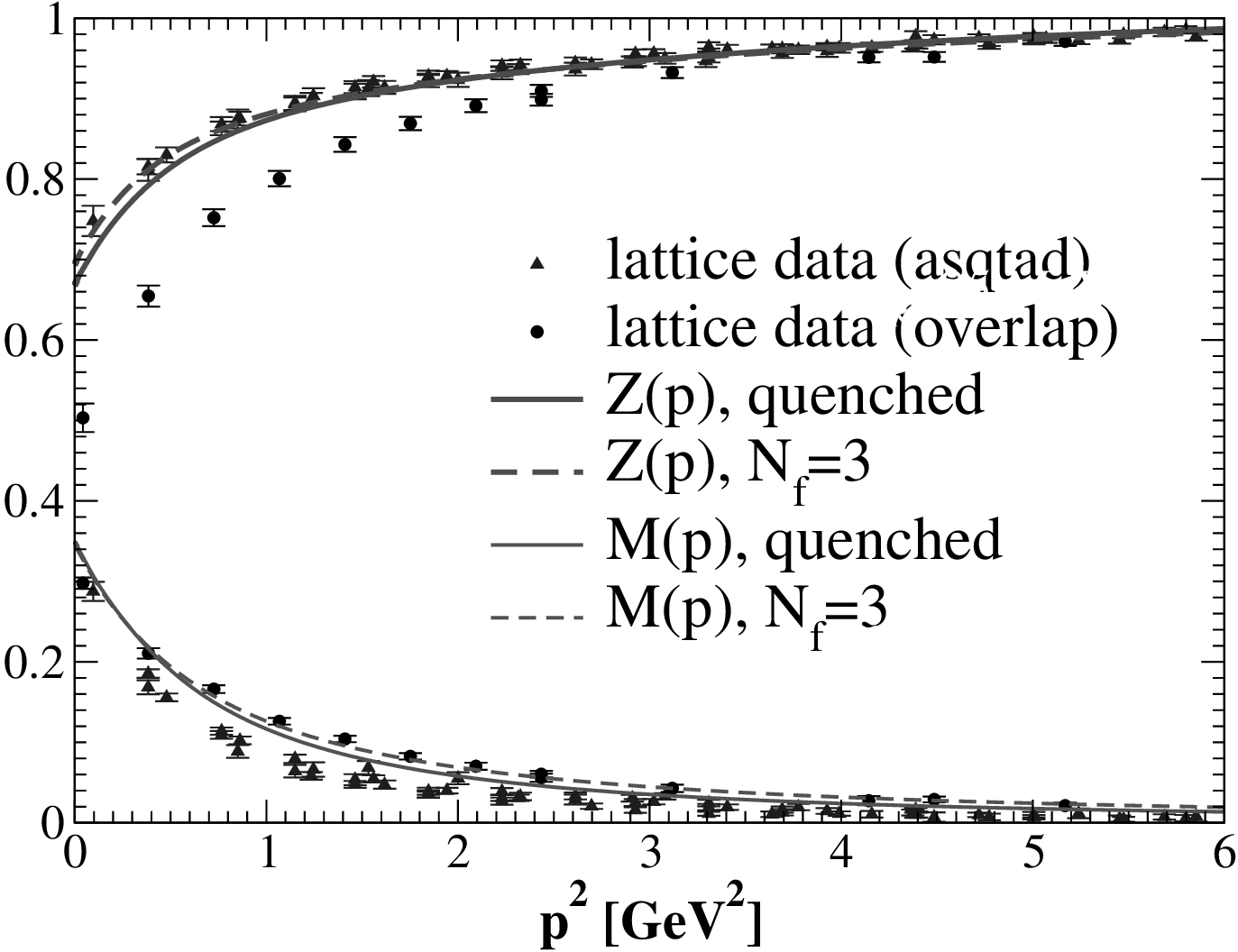}
\vspace{-1cm}
\caption{The quenched and unquenched quark mass function $M(p^2)$ and the 
wave function $Z_Q(p^2)$ from the DSE approach \cite{Fischer:2003rp}
compared to results from quenched lattice calculations \cite{LATTICE_QUARK}.}
\label{fig3}
\end{figure}

The dynamical generation of quark masses can be studied in detail in the
Dyson-Schwinger equation for the quark propagator, Eq.(\ref{quark_prop}).
It is a genuinely non-perturbative phenomenon and requires a 
careful treatment
of the quark-gluon interaction employed. In the DSE-framework of
ref. \cite{Fischer:2003rp} an ansatz for the quark-gluon vertex has been 
chosen that satisfies important constraints: it is genuinely nonperturbative 
in nature, 
guarantees multiplicative renormalisability in the quark DSE and
has the correct limit in the perturbative momentum domain. Its structure is 
such that it factorizes in an Abelian and a non-Abelian part,
\begin{equation}
\Gamma_\nu(q,k) = V_\nu^{Abel}(p,q,k) \, W^{\neg Abel}(p,q,k),
\label{vertex-ansatz}
\end{equation}
with $p$ and $q$ denoting the quark momenta and $k$ the gluon momentum. For 
the Abelian part $V_\nu^{Abel}$ 
carrying all the tensor structure the 
Curtis-Pennington construction \cite{Curtis:1990zs} has been chosen which
has been used frequently in QED.
This choice has the advantage that vector as well as scalar tensor
components are taken into account and their effect on the quark propagator
can be studied. In fact, as detailed in refs. \cite{Alkofer:2003jj},
the inclusion of the scalar parts of the vertex is capable of changing 
the analytic structure of the quark propagator dramatically, leading to 
a positive definite spectral function for the quarks.

The non-Abelian part $W^{\neg Abel}(p,q,k)$ of the quark-gluon vertex 
(\ref{vertex-ansatz})
contains ghost-dressing factors as (partly) implied by its Slavnov-Taylor identity. 
The combined dressing of the quark-gluon vertex and the full gluon propagator 
in the quark DSE then provides enough interaction strength to generate
dynamical quark masses of the order of $M(0) \approx 350$ MeV in agreement
with phenomenology. Furthermore a chiral condensate of the order
$(-\langle \bar{\Psi} \Psi \rangle)^{-1/3} \approx 300$ MeV has been obtained, 
which strongly favours the standard counting rules in chiral perturbation theory.

The results for the quenched quark mass function $M(p^2)$ and the wave function
$Z_Q(p^2)$ are compared to the quenched lattice results of refs.
\cite{LATTICE_QUARK} in fig.~\ref{fig3}. The overall qualitative and quantitative 
agreement between both approaches is very good. The DSE results are within the 
bounds given by the two different formulations of fermions on the lattice. 

\begin{figure}[t!]
\includegraphics[width=\columnwidth]{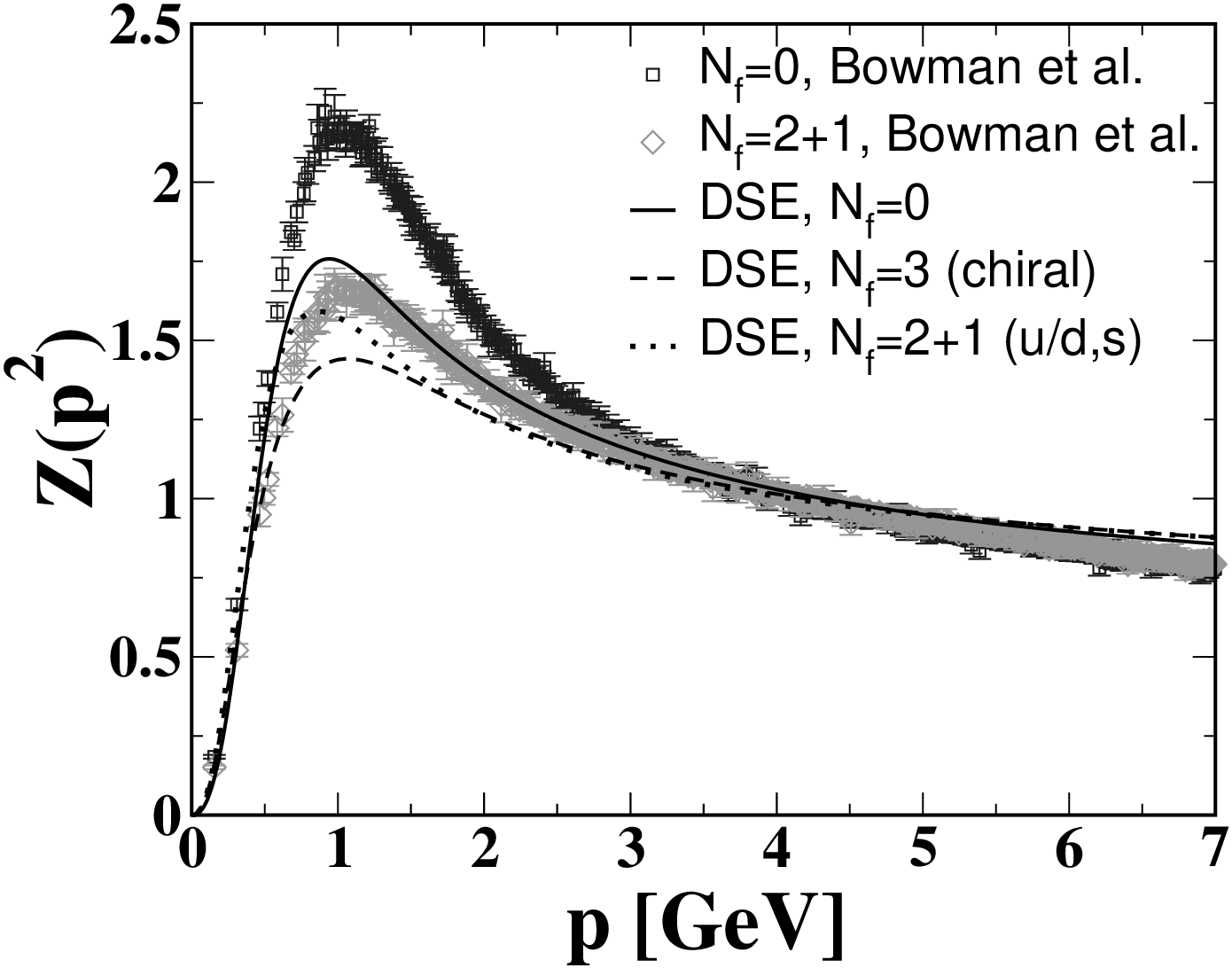}
\vspace{-1cm}
\caption{The quenched and unquenched gluon dressing function from the DSE 
approach \cite{Fischer:2003rp} compared to results from unquenched lattice 
calculations \cite{Bowman:2004jm}.}\label{fig4}
\end{figure}

Including the backreaction of the quark-propagator on the ghost and gluon system
leads to a coupled set of three Dyson-Schwinger equations for the propagators of 
QCD. These equations have been solved in \cite{Fischer:2003rp} and allowed a 
prediction of possible effects of unquenching QCD on the propagators. As can be seen
from fig. \ref{fig3} including $N_f=3$ chiral quarks in the gluon
DSE hardly changes the results for the quark propagator. The slight enhancement
in the perturbative tail of the mass function is in perfect accordance with 
the expected change of the
anomalous dimension. The chiral condensate is nearly unaffected. It will be 
interesting to compare these results to unquenched lattice calculations when
available.

Unquenched lattice results for the gluon propagator including the effects of 
two light (up-) and one heavy (strange-) quark have been published recently
\cite{Bowman:2004jm} and are compared to the corresponding results from our
DSE-approach in fig. \ref{fig4}. The screening effect from the quark loop is
clearly visible in the lattice results for momenta $p$ larger than $p=0.5$ GeV: 
the gluonic self interaction becomes less important in this region and the gluon
dressing increases. Although to a somewhat less extent, this effect can also be 
seen in the DSE-approach. As mentioned already above not all effects from the 
gluonic self interaction are accounted for in the DSE truncation. When this part 
of the gluon interaction becomes less dominant
in the unquenched case, both the lattice and the DSE-approach agree very well on
a quantitative level, provided similar bare quark masses are taken into account.
In the chiral limit the screening effect of the quark loop becomes even stronger
as can be seen from the DSE-results in fig.~\ref{fig4}. This is certainly expected 
as the energy needed to create a quark pair out of the vacuum becomes smaller with
decreasing bare quark mass. 

Both, the lattice calculations and the Green's functions approach agree in the fact 
that unquenching does not affect the extreme infrared of the ghost and gluon 
propagators. Again, this is easily explained from dynamical chiral 
symmetry breaking: there is
not enough energy to generate a quark pair from the vacuum below a certain threshold,
and the quark degrees of freedom decouple from the Yang-Mills sector of the theory.

\section{The quark-gluon vertex \label{sec4}}

\begin{figure}[t!]
\includegraphics[width=\columnwidth,height=5.7cm]{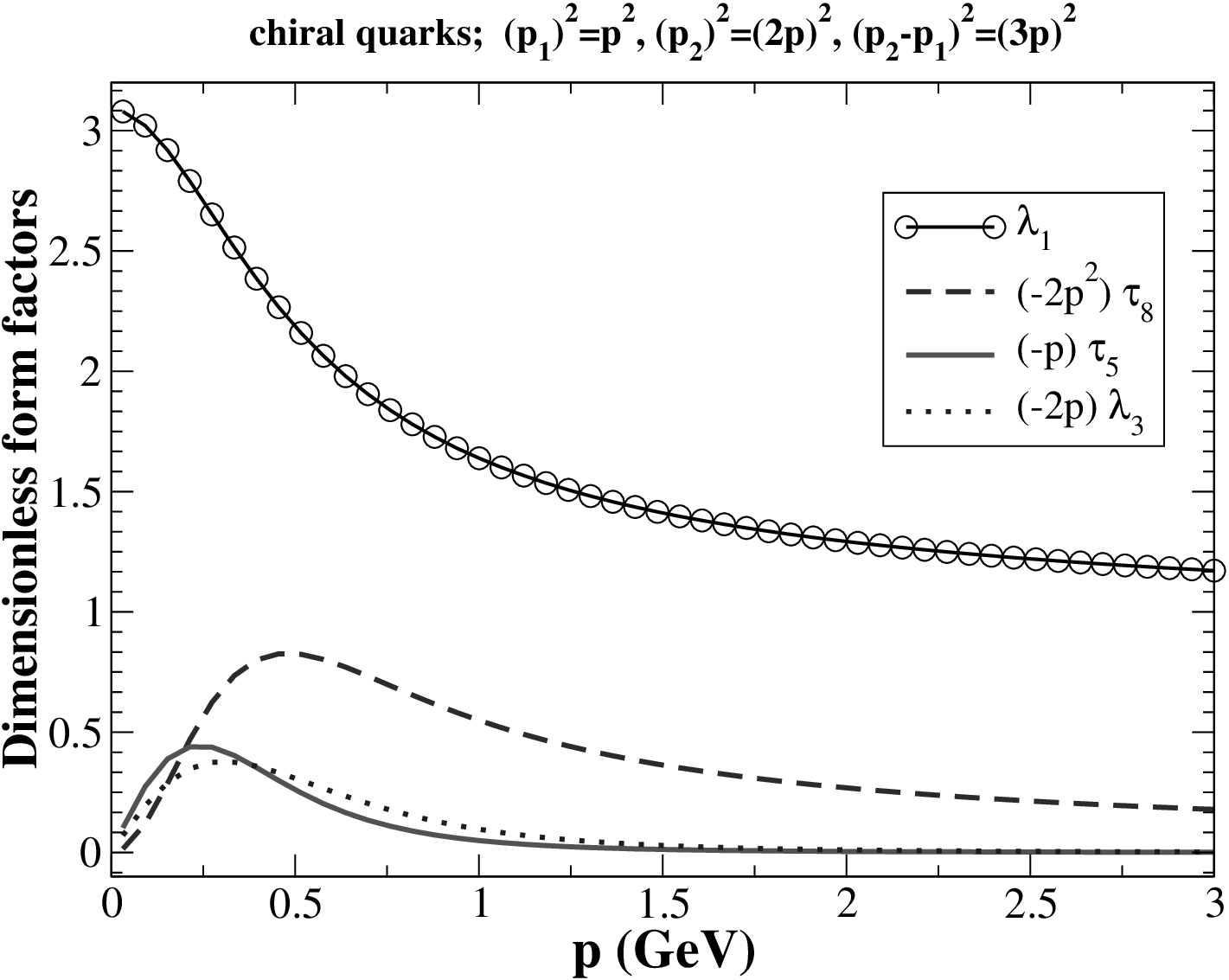}
\vspace{-1cm}
\caption{The leading tensor structures of the quark gluon vertex from the
vertex DSE with chiral quarks.}\label{fig5}
\vspace{-1cm}
\end{figure}

As mentioned in the introduction, the quark-gluon vertex plays an 
important role in both, understanding confinement from the quark-gluon interaction 
and providing the bridge between coloured quarks and gluons, and their colourless 
bound states, the hadrons. Therefore it is desirable to determine its properties 
in a selfconsistent calculation. 
As a first step in this direction we present a calculation of the components
of the quark-gluon vertex from its DSE (also see Ref.\cite{Bhagwat:2004hn}). 
Our calculation is not 
selfconsistent 
yet: 
we use the quark, ghost and gluon propagators presented in the last section as 
input on the right hand side of the vertex DSE without backcoupling the vertex to
the quark equation. Furthermore we only employ the $\gamma_\mu$-part of the vertex 
in all loops of the vertex DSE. Nevertheless we believe that our results are 
meaningful as the propagators used are close to their respective counterparts 
on the lattice. Furthermore from the structure of the vertex DSE one finds
indications
that selfconsistency effects could play a major role in the infrared but are 
small in the medium and large momentum regime.

\begin{figure}[t!]
\includegraphics[width=\columnwidth,height=5.7cm]{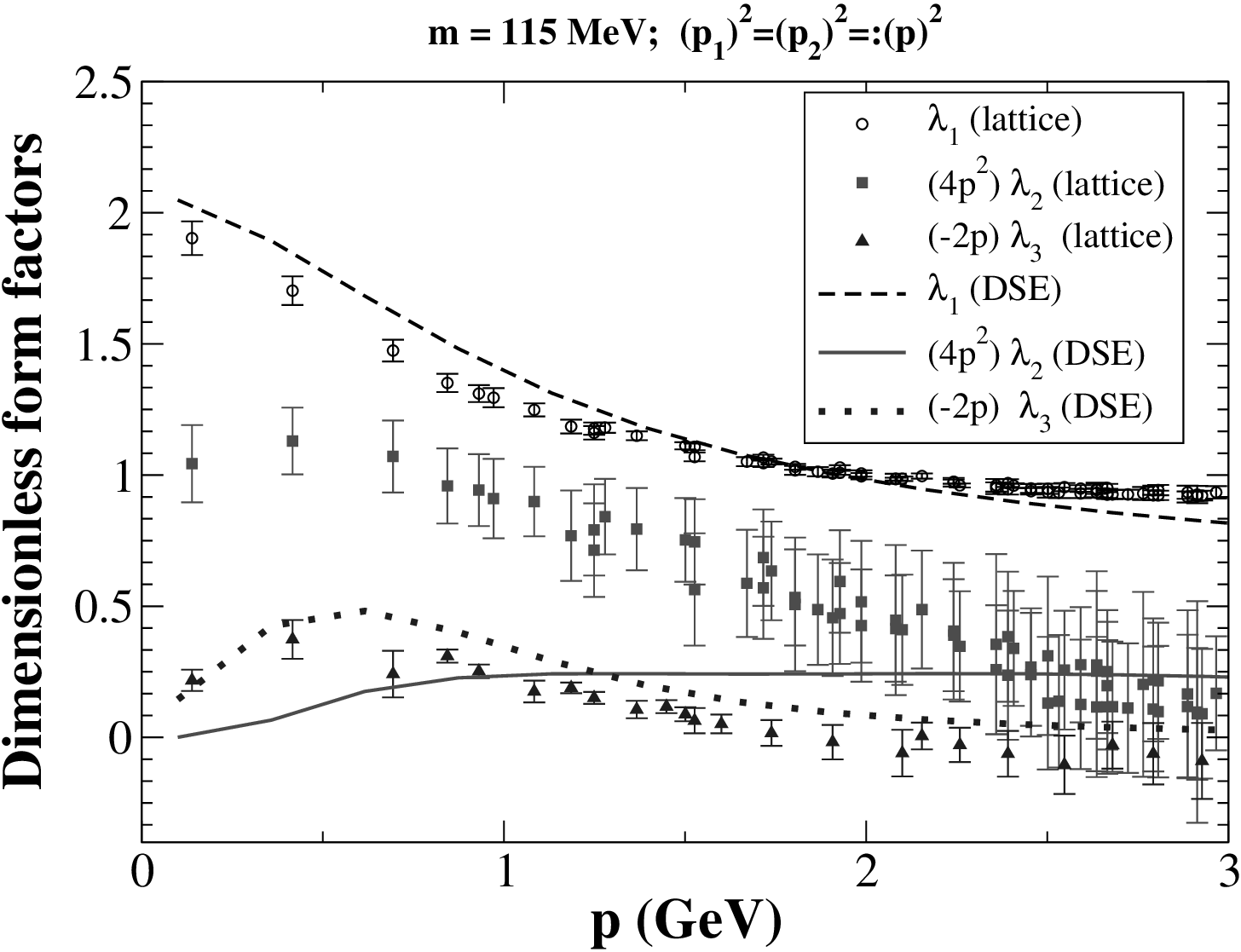}
\vspace{-1cm}
\caption{Tensor structures of the quark gluon vertex from the
vertex DSE with a massive quark compared to lattice calculations 
\cite{Skullerud:2003qu}.}\label{fig6}
\vspace{-0.6cm}
\end{figure}

With these caveats in mind we present our results in figs.~\ref{fig5} and
\ref{fig6}. The complete vertex can be decomposed in a basis of twelve
tensor structures, $\Gamma_\mu = \frac{i}{g}\left(\sum_{i=1}^4 \lambda_i L_{i\ \mu}
+ \sum_{i=1}^8 \tau_i T_{i\ \mu}\right)$\cite{Skullerud:2003qu}. Here we show only 
five of these components,
namely the ones multiplying the tensors 
\beqa \label{newdec}
L_{1\ \mu}&=& \gamma_\mu \nonumber\\ \nonumber
L_{2\ \mu}&=&-(\not p_1 + \not p_2)(p_1+p_2)_\mu \\  \nonumber
L_{3\ \mu}&=& i(p_1+p_2)_\mu \\  \nonumber
T_{5\ \mu}&=& i \sigma_{\mu \nu} (p_2-p_1)^\nu \\  \nonumber
T_{8\ \mu}&=& -\gamma_\mu \sigma_{\lambda \nu} p_1^\lambda p_2^\nu -\not
p_1 p_{2\ \mu} +\not p_2 p_{1\ \mu}
\eeqa
with the quark momenta $p_1$ and $p_2$. The (dimensionless) four leading structures 
in the chiral limit are given in fig.~\ref{fig5}. Although the $L_{1}$-
piece is dominating, sizeable admixtures from other tensor structures occur.
This is also found in a recent model study \cite{Watson:2004kd}
and it is expected, as the quark-gluon vertex provides one of the
underlying structures from which the rich structure of phenomenological 
quark potentials (see e.g. \cite{Ebert:2002pp}) should be ultimately derived.

Our results for a massive quark of $m_0=115$ MeV are compared with the available 
lattice results in fig.~\ref{fig6}. Even without selfconsistency in the DSEs we
obtain very nice agreement for the vector and the scalar tensor 
components, $L_1$ and $L_3$. On the other hand, the infrared behaviour of the 
$L_2$-piece is in obvious disagreement.
Firm conclusions, however, cannot be drawn before lattice data with smaller
errors as well  as a selfconsistent DSE  solution are available.

\newpage
\section{Outlook}

In the last years a consistent picture of the infrared behaviour of Landau 
gauge 
QCD has emerged. Evidence from the Green's functions approach and lattice 
calculations suggests that the Faddeev-Popov determinant (i.e. the ghost 
degrees of freedom) is dominant in the infrared and provide the long range 
interaction in QCD. The gluon propagator is finite or even vanishing in the 
infrared. Taken at face value this behaviour provides a problem for 
confinement: Simple one-gluon exchange with a bare quark-gluon vertex
cannot account for the linear rising potential between two static colour 
sources. Infrared singularities present in the dressed quark-gluon vertex
could resolve this issue. Whether this is indeed the case or a more
sophisticated mechanism is at work remains an open problem for the future.

\section*{Acknowledgments}

We are grateful to M.~Bhagwat, H.~Gies, C.~Roberts, J.~Skullerud,
P.~Tandy, A.~Williams, and D.~Zwanziger for helpful discussions.  

This work has been supported by the Deutsche For\-schungsgemeinschaft
(DFG) under contracts Al 279/3-3, Al 279/3-4, Fi 970/2-1 and GRK683.
F.L. thanks the German DAAD and Univ. Complutense for support and 
the members of the Institute for Theoretical Physics 
of the University of T\"ubingen for their hospitality during his visit,
when the project reported in section \ref{sec4} has been started. 

We thank A.\ Kizilers{\"u}, A.\ Thomas and A.\ Williams for
the organisation of this exceptional conference QCD DOWN UNDER.


\begin{thebibliography}{99}
\parskip=0pt

\bibitem{Maris:2003vk}
P.~Maris and C.~D.~Roberts,
Int.\ J.\ Mod.\ Phys.\ E {\bf 12}, 297 (2003). 

\bibitem{Alkofer:2000wg}
R.~Alkofer and L.~von Smekal,
Phys.\ Rept.\  {\bf 353}, 281 (2001).

\bibitem{Roberts:2000aa}
C.~D.~Roberts and S.~M.~Schmidt,
Prog.\ Part.\ Nucl.\ Phys.\  {\bf 45}, S1 (2000).

\bibitem{vonSmekal:1997is}
L.~von Smekal, R.~Alkofer and A.~Hauck,
Phys.\ Rev.\ Lett.\  {\bf 79}, 3591 (1997); 
Annals Phys.\  {\bf 267}, 1 (1998).

\bibitem{Fischer:2002hn}
C.~S.~Fischer  and R.~Alkofer,
Phys. Lett. B {\bf 536}, 177 (2002);
R.~Alkofer, C.~S.~Fischer and L.~von Smekal,
Acta Phys.\ Slov.\  {\bf 52}, 191 (2002).

\bibitem{Fischer:2002eq}
C.~S.~Fischer, R.~Alkofer and H.~Reinhardt,
Phys.\ Rev.\ D {\bf 65}, 094008 (2002).

\bibitem{Lerche:2002ep}
C.~Lerche and L.~von Smekal,
Phys.\ Rev.\ D {\bf 65}, 125006 (2002).

\bibitem{Zwanziger:2001kw}
D.~Zwanziger,
Phys.\ Rev.\ D {\bf 65},  094039 (2002).

\bibitem{Watson:2001yv}
P.~Watson and R.~Alkofer,
Phys.\ Rev.\ Lett.\  {\bf 86}, 5239 (2001).

\bibitem{Nakanishi:qm}
N.~Nakanishi and I.~Ojima,
``Covariant Operator Formalism Of Gauge Theories And Quantum Gravity,''
World Sci.\ Lect.\ Notes Phys.\  {\bf 27}, 1 (1990).

\bibitem{Kugo:1979gm}
T.~Kugo and I.~Ojima, Prog.~Theor.~Phys.~Suppl. {\bf 66}, 1 (1979).

\bibitem{Zwanziger:2003cf}
D.~Zwanziger,
Phys.\ Rev.\ D {\bf 69}, 016002 (2004).

\bibitem{Pawlowski:2003hq}
J.~M.~Pawlowski, D.~F.~Litim, S.~Nedelko and L.~von Smekal,
arXiv:hep-th/0312324;
C.~S.~Fischer and H.~Gies, {\it in preparation}.


\bibitem{Fischer:2003rp}
C.~S.~Fischer and R.~Alkofer,
Phys.\ Rev.\ D {\bf 67}, 094020 (2003).

\bibitem{Shirkov:1997wi}
D.~V.~Shirkov and I.~L.~Solovtsov,
Phys.\ Rev.\ Lett.\  {\bf 79}, 1209 (1997).

\bibitem{LATTICE_QUARK}
P.~O.~Bowman, U.~M.~Heller and A.~G.~Williams,
Phys.\ Rev.\ D {\bf 66}, 014505 (2002);
J.~B.~Zhang {\it et al.\/}, 
arXiv: hep-lat/0301018.







\bibitem{Bonnet:2001uh}
F.~D.~Bonnet {\it et al.},
Phys.\ Rev.\ D {\bf 64}, 034501 (2001).

\bibitem{Bowman:2004jm}
P.~O.~Bowman {\it et al.\/},
arXiv:hep-lat/0402032.



\bibitem{Curtis:1990zs}
D.~C.~Curtis and M.~R.~Pennington,
Phys.\ Rev.\ D {\bf 42}, 4165 (1990).

\bibitem{Alkofer:2003jj}
R.~Alkofer, W.~Detmold, C.~S.~Fischer and P.~Maris,
arXiv:hep-ph/0309077;
R.~Alkofer, W.~Detmold, C.~S.~Fischer and P.~Maris,
arXiv:hep-ph/0309078.

\bibitem{Bhagwat:2004hn}
M.~S.~Bhagwat {\it et al.}, 
arXiv:nucl-th/0403012.

\bibitem{Skullerud:2003qu}
J.~I.~Skullerud {\it et al.}, 
JHEP {\bf 0304} (2003) 047.

\bibitem{Watson:2004kd}
P.~Watson, W.~Cassing and P.~C.~Tandy,
arXiv:hep-ph/0406340.


\bibitem{Ebert:2002pp}
D.~Ebert, R.~N.~Faustov and V.~O.~Galkin,
Phys.\ Rev.\ D {\bf 67} (2003) 014027.



\end{thebibliography}
\end{document}